\documentclass[prb,aps,twocolumn,amsfonts]{revtex4}

\usepackage{amsmath}      
\usepackage{graphicx}     
\usepackage{verbatim}     
\usepackage{color}        

\let\omp\marginpar\relax
\def\marginpar#1{\omp{\color{red}#1}}
\newlength\figurewidth
\AtBeginDocument{\setlength\figurewidth{0.98\linewidth}}

\def\Ang{\AA}
\def\Vshear{\AA/ps}
\def\aSi{{\it a}-Si}
\def\eal{\emph{et al.}}

\begin{document}

\title{Amorphous silicon under mechanical shear deformations: shear velocity and temperature effects}

\author{Ali Kerrache} \email{ali.kerrache@umontreal.ca}
\affiliation{D\'epartement de Physique, Regroupement Qu\'eb\'ecois sur les Mat\'eriaux de Pointe, Universit\'e de Montr\'eal, C.P. 6128, succ. Centre-ville, Montr\'eal (Qu\'ebec) H3C 3J7, Canada.}

\author{Normand Mousseau} \email{normand.mousseau@umontreal.ca}
\affiliation{D\'epartement de Physique, Regroupement Qu\'eb\'ecois sur les Mat\'eriaux de Pointe, Universit\'e de Montr\'eal, C.P. 6128, succ. Centre-ville, Montr\'eal (Qu\'ebec) H3C 3J7, Canada.}

\author{Laurent J. Lewis} \email{laurent.lewis@umontreal.ca}
\affiliation{D\'epartement de Physique, Regroupement Qu\'eb\'ecois sur les Mat\'eriaux de Pointe, Universit\'e de Montr\'eal, C.P. 6128, succ. Centre-ville, Montr\'eal (Qu\'ebec) H3C 3J7, Canada.}

\date{\today}

\begin{abstract}

Mechanical shear deformations lead, in some cases, to effects similar to those resulting from ion irradiation. Here we characterize the effects of shear velocity and temperature on amorphous silicon (\aSi) modelled using classical molecular dynamics simulations based on the empirical Environment Dependent Inter-atomic Potential (EDIP). With increasing shear velocity at low temperature, we find a systematic increase in the internal strain leading to the rapid appearance of structural defects (5-fold coordinated atoms). The impacts of externally applied strain can be almost fully compensated by increasing the temperature, allowing the system to respond more rapidly to the deformation. In particular, we find opposite power-law relations between the temperature and the shear velocity and the deformation energy. The spatial distribution of defects is also found to strongly depend on temperature and strain velocity. For low temperature or high shear velocity, defects are concentrated in a few atomic layers near the center of the cell while, with increasing temperature or decreasing shear velocity, they spread slowly throughout the full simulation cell. This complex behavior can be related to the structure of the energy landscape and the existence of a continuous energy-barrier distribution.

\end{abstract}


\maketitle

\section{\label{intro}Introduction}

The structural and dynamical properties of amorphous silicon (\aSi) have been extensively investigated using classical force fields~\cite{WWW, Mousseau98, Barkema98, Balamane, Bazant, Justo, SW, Tersoff, Gillespie, Kluge, Biswas, Valiquette}, tight-binding~\cite{Bernstein06, Makhov, Feldman, Huang} and \emph{ab-initio} simulations~\cite{Drabold, Stich, Car}. Much attention, in particular, has been devoted to the characterization of defects~\cite{Valiquette, Mousseau00, David, Roorda, Coffa, Raymond, Knief, Urli, Peressi, Goedecker, Amarendra, Pantelides, Kim}. In spite of these efforts, the nature and role of defects in disordered materials is not fully understood and numerous questions remain, especially with regards to relaxation. For example, the usual definition of defects --- such as vacancies and interstitials --- cannot be directly applied to disordered or amorphous materials: while 3- and 5-fold coordinated atoms are often taken as defect centers in \aSi, other medium-range defects, such as string of \emph{short} or \emph{long} bonds, might also have to be considered, as shown recently by Drabold and collaborators~\cite{zhang08, Drabold10}.

We study here the role of defects with respect to relaxation by examining the response of \aSi~ to plastic deformations. In crystalline materials, where the phenomenon is well understood, plasticity is attributed to defect nucleation or dislocation motion~\cite{Rice, Acharya, Gaucherin, Mesarovic, Bonifaz, Devincre, Enikeev, Hwang, Gao, Han}; in disordered materials, the response is harder to define structurally, as demonstrated by a number of studies in systems ranging from metallic~\cite{Falk04} and polymeric~\cite{Ediger} glasses to granular materials~\cite{Liu93} and colloids~\cite{Weeks}. Helder \eal~\cite{Helder}, however, have shown that, during irradiation with high-energy heavy ions, \aSi~ deforms plastically in the same manner as conventional glasses, i.e., defects, irradiation and plasticity are directly related in amorphous materials. It is this still imperfectly understood relation that motivates the work presented here.

Following initial suggestions by Argon~\cite{ArgonSTZ}, it has been proposed recently that plasticity is caused by collections of shear transformation zones that operate as localized centers for the deformations.~\cite{Falk, Falk04, Langer06,Argon, Demkowicz}  In \aSi, these plastic deformations are attributed to the presence of liquid-like particles~\cite{Argon, Demkowicz} associated to 5-fold coordinated atoms. To verify these ideas, elastic and plastic deformations in \aSi~ were investigated by Talati~\eal~\cite{Talati} using classical MD simulations based on the Stillinger-Weber~\cite{SW} and the Tersoff~\cite{Tersoff} potentials; it was concluded that, even though the general behavior of the stress-strain curves associated with elastic and plastic deformations are similar to that for other disordered materials, details as to the nature of the defect responsible for plasticity depend on the particular potential used.

We revisit this question here using classical MD simulations and the Environment Dependent Interatomic Potential (EDIP)~\cite{Bazant, Justo}. Various points defects and their effects on elastic constants have been characterized by Allred~\eal~\cite{Allred} using the EDIP potential~\cite{Bazant, Justo}. These authors have shown that the elastic constants vary in a roughly linear fashion with defect concentration up to $\sim$3\%, in line with experiment, suggesting that EDIP is suitable for investigating the elastic and plastic deformations in \aSi. Our simulations were performed using the classical MD package (LAMMPS)~\cite{LAMMPS}.

Most previous studies of elastic and plastic deformations in amorphous materials have been performed at 0~K or at low temperature, with a focus on the disordering process. Here we examine the situation in \aSi~by investigating the system's response to variations in shear velocity and temperature, as both parameters contribute, in their own way, in forcing the system to overcome energy barriers and explore the potential energy surface; as mentioned previously, structural changes are strongly correlated with imposed strain~\cite{Talati, Ivashchenko}.

This paper is organized as follows: in the section~\ref{model}, we describe the details of our simulation model. In the next section (\ref{res}), we present the results obtained by applying mechanical shear deformations at different shear velocities and temperatures. In the section~\ref{structure}, we present the structural analysis as a function of shear velocity. After discussion (section~\ref{discussion}) of our results, we will give our main conclusions (section~\ref{conclude}).

\section{\label{model}Computer simulation model}

Classical MD simulations have limitations, but they are unavoidable to simulate systems large enough to limit the impact of size effects. For the problem at hand, charges and electronic effect should not play a direct role and therefore a description in terms of suitable classical interactions is appropriate.

The EDIP functional form involves a two-body radial term for bond-stretching interactions and a three-body angular term for bond-bending interactions, and each of these depends strongly on an effective coordination number $Z$. We have chosen this potential for its ability to reproduce a wide range of zero-temperature properties of Si, including elastic constants, bulk crystal structures and point defects~\cite{Justo, Allred} --- in particular, EDIP describes accurately the vacancy formation energy. This potential has been used to study the ion-beam induced amorphization of crystalline silicon~\cite{Caturla, Nord} and the crystallization~\cite{Nakhmanson, Pelaz, Park} of \aSi~starting from amorphous-crystalline interfaces~\cite{Krzeminski}.

The melting temperature of crystalline silicon obtained using EDIP is 1500~K and it is 1200~K for \aSi. In both cases, this is roughly 200~K below experimental values (1685~K and 1420~K, respectively~\cite{Donovan}). Both phases are therefore stable or, at least, metastable in the temperature range investigated here.

\subsection{\label{preparation}Sample preparation}

Experimentally, \aSi~ can be obtained by laser-melting and quenching, by chemical vapor deposition~\cite{Matsumura}, or by
ion-irradiation~\cite{Laaziri}. Numerically, our models of \aSi~are generated using the modified Wooten-Winer-Weaire (WWW) bond-switching algorithm, which can produce perfect 4-fold-coordinated random networks~\cite{WWW, Barkema} in good structural and electronic agreement with experiments~\cite{Barkema}. A 1000-atom \aSi~ cell was first constructed, then duplicated in order to obtain a cubic box containing 8000 atoms. The latter was then annealed in the NPT ensemble with periodic boundary conditions over 150~ns at 300~K so as to obtain a well relaxed models of amorphous silicon. The final structure was found to contain less than 3\% of defects (mostly 5-fold coordinated atoms), the rest being perfectly 4-fold coordinated, in agreement with the best finite-temperature models available in the literature.

\subsection{\label{procedure}Procedure for mechanical shear deformations}

\begin{figure}[h!]
\includegraphics[width=0.92\figurewidth]{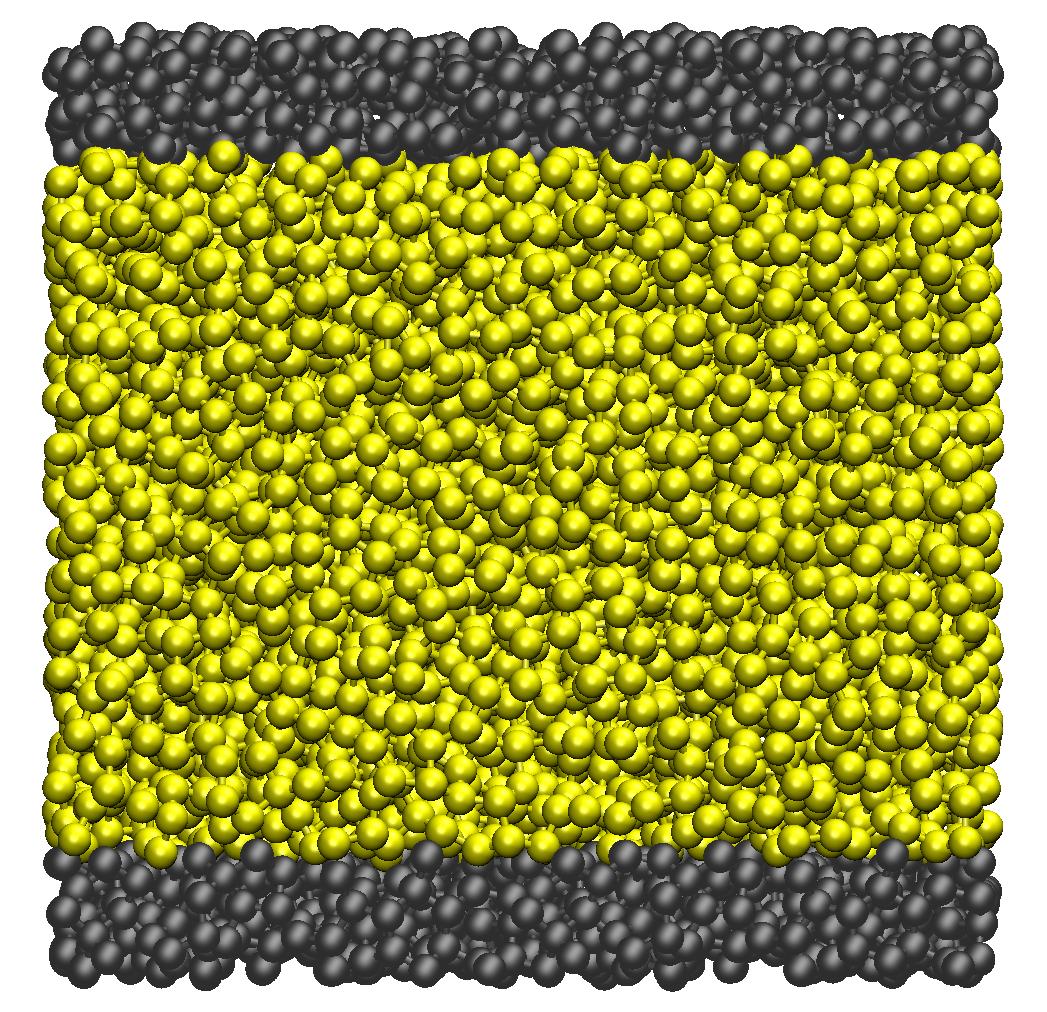}

\caption{Snapshot of a typical configuration of the 8000-particle \aSi-model. Mobile particles are positioned between two parallel walls, each containing 1000 atoms, used for applying the shear deformations (see text for more details).}
\label{FigSnapShear}
\end{figure}

The mechanical shear deformations on the 8000-atom \aSi~model are introduced as follows: the cell is first equilibrated (at the target temperature --- see below) with periodic boundary conditions in all three directions, for at least 5~ns. Three different regions are then defined, as shown in Fig.~\ref{FigSnapShear}: an upper and a lower ``wall'', each containing 1000 atoms, and a region of mobile or active particles; periodic boundary conditions are now imposed only in the lateral ($x$ and $z$) directions, while surface atoms are fixed in the $y$-direction. This procedure is similar to that used by Mokshin~\eal~\cite{Mokshin} on single-component Lennard-Jones amorphous systems. The thickness of the walls, about 6~\Ang, is larger than the EDIP cutoff~\cite{Bazant, Justo}, thus ensuring that all particles within the active region (40~\Ang-thick) share the same physics.

The mechanical shear deformations are generated by moving the walls at fixed shear velocity (shear rate) $v_s$ along the shear direction ($x$). The shear velocity ($v_s$) measures the speed of the deformation while the strain rate ($\dot{\epsilon}$) gives the change in strain with respect to time and it corresponds to the shear velocity devided by the distance between the walls. In practice, the lower-wall position is fixed and only the particles in the upper wall are moved by imposing a constant displacement $v_s \times h$ at every timestep $h$ (1~fs) --- the wall particles are otherwise frozen in place. The amorphous nature of the walls ensures that crystal growth, if it occurs, is not induced by the boundaries as in the case of crystallization studies using amorphous-crystalline interfaces~\cite{Buta, Brambilla, Bernstein98, Mattoni}. Note that the cell is aged during 1~ns before applying the deformations in order to ensure proper relaxation after releasing the periodic boundary conditions along $y$.

The equations of motion for mobile particles are integrated using the velocity Verlet algorithm. Since an additional force is imposed in the shear direction ($x$), only the components in the ``neutral'' $y$ and $z$ directions are considered for computing and rescaling the velocities to constant temperature during the deformation process. The pressure at time $t$ is computed from instantaneous atomic positions [$\overrightarrow{r_i}(t)$, $i=1,N$] and forces acting on the particles [$\overrightarrow{F_i}(t)$, $i=1,N$] using $P(t) = \frac{N}{V}\, k_BT + \frac{1}{3V} \sum_{i=1}^{N}\overrightarrow{r_i}(t) \cdot \overrightarrow{F_i}(t)$, with $N$ the number of particles, $V$ the volume, $T$ the temperature, and $k_B$ Boltzmann's constant. Statistics such as potential energy and coordination number were obtained at zero shear velocity and will be used as reference for assessing the effect of the shear deformations.

The elastic and plastic deformations of materials are usually analyzed in terms of stress-strain curves~\cite{Ivashchenko, Talati, Rottler} where the strain corresponds to the maximum displacement in the shear direction with respect to the distance between the two walls, which is kept constant. We prefer to characterize the deformations in terms of the potential energy difference (PED) $\Delta E = E_p - E_0\,$, where $E_p\,$ and $E_0\,$ are the potential energies of sheared and non-sheared systems; $E_0\,$ is computed before switching on the shear deformations. This analysis has the advantage of providing direct microscopic information and relates more readily to structural changes during shearing. Structural properties like radial distribution functions and the coordination number will also be used to determine how the properties of \aSi~are affected by mechanical shear deformations.

\section{\label{res}Shear deformations results}

In this section, we examine the effects of varying the shear velocity and the temperature on the system's structural disorder and defect concentration.

\subsection{Mechanical shear deformations at 300~K}
\label{res300}

We fix the temperature at 300~K and vary $v_s$ between $10^{-5}$ and $8 \times 10^{-2}$~\Vshear. For all values of $v_s$ but the smallest, simulations are run until a strain of 20\% is reached; e.g., for a shear velocity of $10^{-2}$~\Vshear, the simulation time is 0.8~ns. For the lowest shear velocity ($v_s=10^{-5}$~\Vshear), because of computational limitations, we stopped at 12\% strain, corresponding to a simulation time of 500~ns. In all cases, simulations are long enough for the plastic deformation threshold --- defined by the onset of irreversible deformations --- to be reached; this occurs in \aSi~ at a strain of about 10\%. The exact value of this threshold depends on the shear velocity and we found it to increase with $v_s$, in agreement with previous results (see Ref.~\cite{Ivashchenko}, for example).

It is well known that stress appears immediately following a shear deformation. At short times or small strain values, the stress increases almost linearly and saturates at high strain values. This constant stress corresponds to the sheared steady-state character of plastic deformations. This general behavior is common for polymeric materials~\cite{Ediger}, metallic glasses~\cite{Falk04}, colloids~\cite{Weeks} and amorphous materials~\cite{Helder, Argon, Demkowicz, Ivashchenko, Talati}. The response of the system to mechanical shear deformations can be analyzed, in general, in terms of stress-strain curves~\cite{Ivashchenko, Talati}. As mentioned above, we prefer to use the PED between sheared and non-sheared cells, as it provides a direct and simple description of the microscopic structural deformations that take place under shear deformations.

\begin{figure}[h!]
\includegraphics[width=0.9\figurewidth]{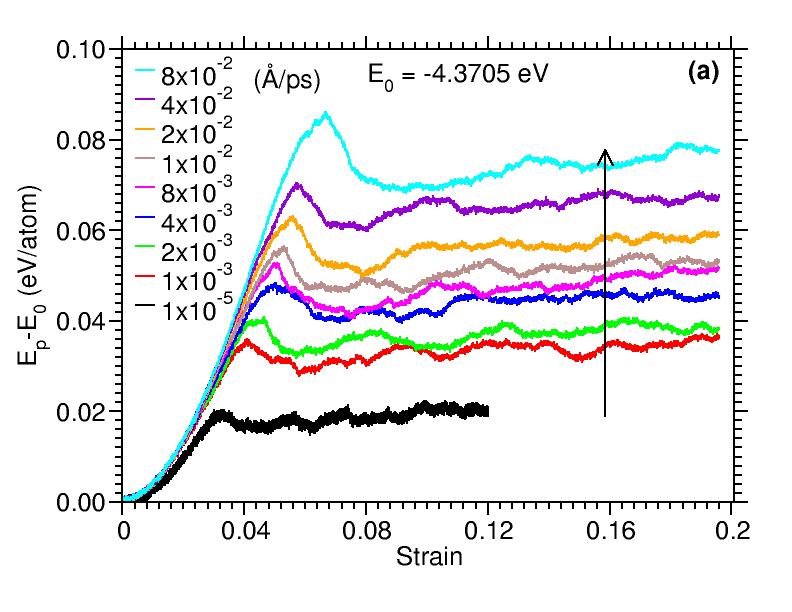}
\includegraphics[width=0.9\figurewidth]{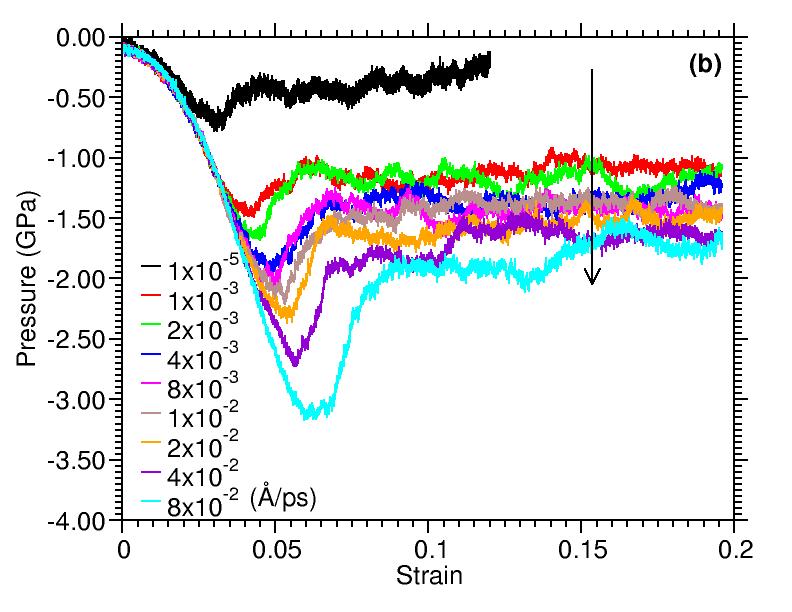}
\includegraphics[width=0.9\figurewidth]{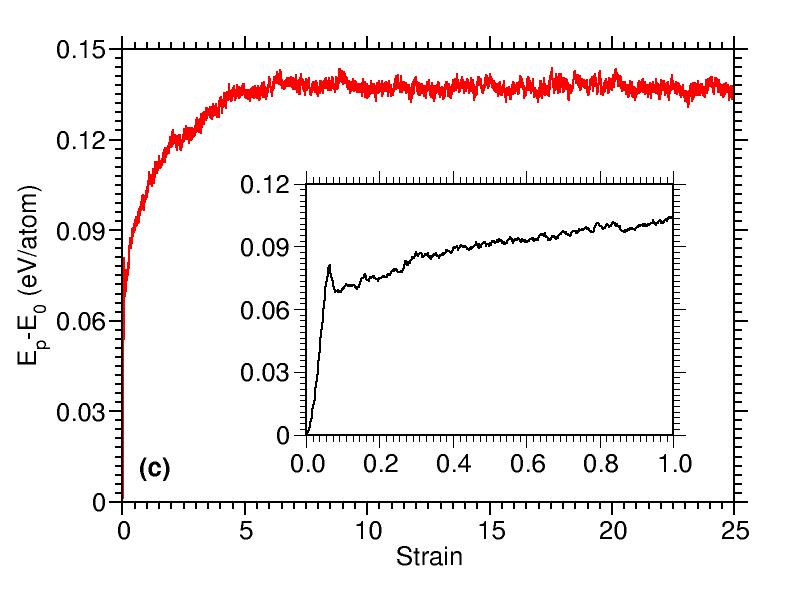}

\caption{(a) Potential energy difference and (b) pressure difference between sheared and non sheared system at 300~K as a function of strain, for shear velocities in the range $10^{-5}$ to $8 \times 10^{-2}$~\Vshear. The potential energy of the non-sheared system ($-4.3707$~eV/atom) is used as reference to calculate the PED. The arrows refer to the direction of increasing shear velocities. (c) Potential energy difference for higher strain values using a shear velocity of $8 \times 10^{-2}$~\Vshear. The inset zooms on the low-strain values.}

\label{FigEpotPressShear300}
\end{figure}

Figure~\ref{FigEpotPressShear300} shows (a) $\Delta E$, (b) the internal pressure as a function of strain for different shear velocities and (c) $\Delta E$ higher strain values using a shear velocity of $8 \times 10^{-2}$~\Vshear (Arrows indicate the direction of increasing shear velocity). Evidently, changes in the potential energy are a manifestation of structural rearrangements that have taken place with respect to the initial, unstrained model. Thus, mechanical shear --- and therefore strain --- increases the potential energy, i.e., the disorder, of the system (cf.\ also Ivashchenko~\eal~\cite{Ivashchenko}).

Changes in the potential energy as a function of shear take place over two different regimes. For small shears, a quadratic behaviour is observed, associated with elastic and reversible deformations; the quadratic nature of the energy is independent of the rate at which the strain is applied, as one can see from the overlap of the curves corresponding to different shear velocities. The point at which the system crosses over to the second, high-strain regime, however, does depend on $v_s$; the larger $v_s$, the longer the elastic regime persists. This behavior is perfectly echoed in the pressure --- a larger $v_s$ allows the system to reach a much higher negative pressure before plastic deformations are forced. The potential-energy maximum also corresponds to a stress maximum in the stress-strain curves or the yield stress (not shown). This behavior is common for elastic and plastic deformations of polymeric~\cite{Ediger}, metallic glasses~\cite{Falk04}, and amorphous materials~\cite{Rottler}.

The onset of plastic deformations is characterized by an overall relaxation of the system, as we observe in Fig.~\ref{FigEpotPressShear300}(a). The higher the yield stress (defined by the pressure, for example), the larger the relaxation; for the smallest $v_s$, $10^{-5}$~\Vshear, for example, there is very little relaxation after the plastic deformation peak is reached. We note however that, after an initial drop, the potential energy increases again, but at a relatively slow rate. It would eventually saturate to a steady-state level, beyond the reach of our simulations, which corresponds to the plasticity regime associated with a flowing steady state, as one can see in Fig.~\ref{FigEpotPressShear300}(c). In this simulation, performed with a significantly faster shear velocity, the potential energy saturates within the run time indicating that the system has reached the flowing steady state and therefore its maximum disorder. The inset of this figure shows the details for strain values up to 1.0. For all shear velocities considered here, the mean value of the potential energy in the plastic steady state is higher than its value before shear deformations are applied. In other terms, the system exhibits an irreversible plastic deformation leading to increased internal strain as in the case of metallic and silica glasses under high energy irradiation~\cite{Klaumunzer, Hou}. As for irradiation, the mechanical shear deformations lead to the formation of new defects in \aSi; this will be discussed in the section \ref{defectssec}.

In the case of plastic deformations, the mean value of the steady-state potential energy increases with $v_s$, as can be seen in Fig.~\ref{FigEpotPressShear300}(a) --- the curves for different values of $v_s$ are almost parallel, shifted to lower energies for smaller shear velocities. In all cases, we observe a drop in the potential energy after the crossover from elastic to plastic deformations. This suggests that, at this point, the system has to overcome an energy barrier to launch a relaxation cascade. The height of this barrier decreases with increasing shear, and the probability of crossing it at lower strain increases for systems under slower shear rates, explaining the observed behavior. Because of the Boltzmann factor governing the jump rate, the transition is also be pushed to lower strain values with increasing temperature. For example, the minimum strain value is found to lie between $\sim$2\% for the lowest shear velocity and $\sim$8\% for the highest in our 300~K runs while, according to Talati~\eal~\cite{Talati}, it can reach 20\% at 0~K. Clearly, temperature plays an important role on the yield-stress value.

At 300~K, the PED between sheared and annealed \aSi~is about 0.02~eV/atom for $v_s=10^{-5}$~\Vshear~and 0.08~eV/atom for $v_s=8\times10^{-2}$~\Vshear: evidently, the structural changes taking place in \aSi~are strongly correlated to the applied shear velocity. While it was not possible to apply a slower strain rate, these results suggest that the steady-state PED from the annealed state would tend to zero as the strain rate decreases. This question will be discussed in more details in Section \ref{discussion}.

The shear-induced disorder leads to the formation of higher-density regions associated with higher-coordinated, ``liquid-like'' particles. This would explain the increase in amplitude of the negative pressure observed in Fig.~\ref{FigEpotPressShear300}(b) for different shear velocities. For very small strain, the pressure decreases harmonically and reaches a minimum before settling to a steady-state value
associated with the plastic-deformation regime, which is strain independent (as in the case of the PED). For larger strain, the pressure does depend on the shear velocity: the change is 2~GPa for $v_s=8\times 10^{-2}$~\Vshear, while it is only 1~GPa for $v_s=10^{-3}$~\Vshear. The plastic deformations are accompanied by a noticeable drop in the pressure amplitude ~\cite{Bulatov}.This behaviour does not appear to be caused solely by the unusual phase diagram of silicon, where the liquid is denser than the solid, as it is also observed in metallic glasses and is rather due the concentration of defects observed in the plastic regime (see below).

\subsection{\label{restemp}Deformations at fixed shear velocity: temperature effects}

In the previous section, we saw that a higher shear velocity leads to a larger deformation energy in the plastic deformation regime at a given temperature. In this section, we discuss the role of temperature on the elastic and plastic properties of \aSi~ at fixed shear velocity. Very few such studies have been reported on disordered systems and most studies on \aSi~were carried out at low temperature --- 0 or 300~K~\cite{Maloney}. The effects of temperature and shear velocity in amorphous polymers and Lennard-Jones glasses were investigated by Rottler~\eal~\cite{Rottler} who found that they are akin to redefining the time scale for structural modifications: these parameters modify the properties of the glassy state by altering the aging process and inducing rejuvenation. At low temperature, atomic diffusion in glassy materials is completely local and negligible. Increasing the temperature or the strain facilitates the possibility for particles to escape from the cage formed by surrounding particles, thus accelerating diffusion and relaxation.

\begin{figure}[h!]
\includegraphics[width=0.92\figurewidth]{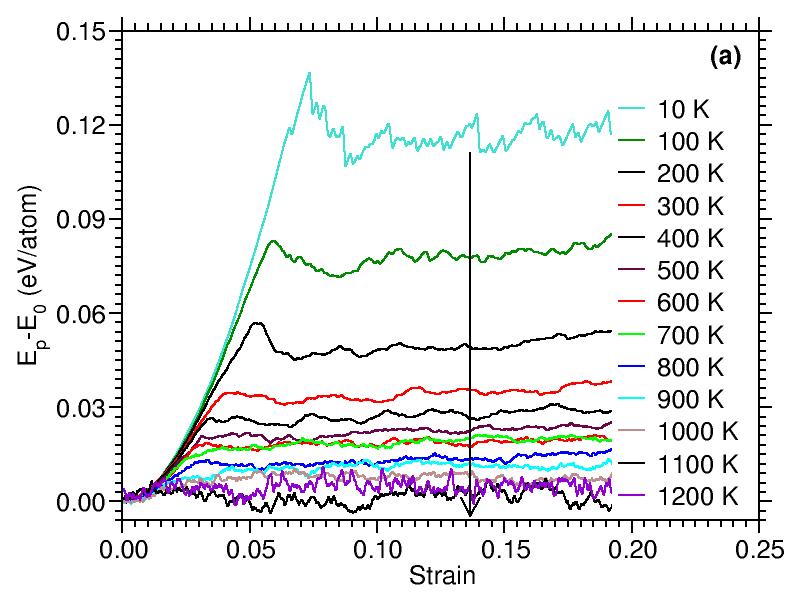}
\includegraphics[width=0.92\figurewidth]{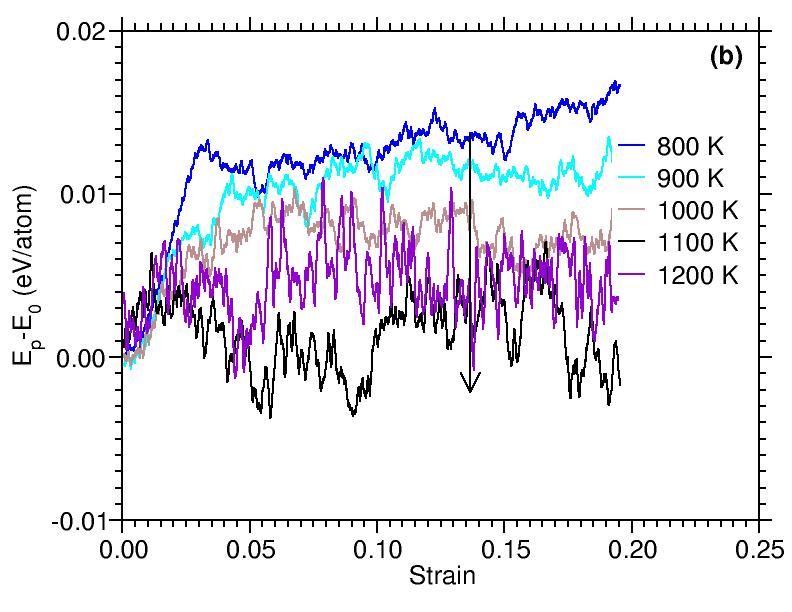}

\caption{(a) Potential energy difference (PED) as a function of strain at fixed shear velocity $v_s = 10^{-3}$~\Vshear~for temperatures (from top to bottom) 10~K, 100~K, 200~K, 300~K, 400~K, 500~K, 600~K, 700~K, 800~K, 900~K, 1000~K, 1100~K and 1200~K. For clarity, the high-temperature curves (800-1200~K) are reproduced in (b) on a finer scale.}

\label{FigEpotTemp}
\end{figure}

In order to isolate the specific effects of temperature, we varied the latter between 10 and 1200~K at fixed shear velocity $v_s=10^{-3}$~\Vshear, during 8~ns for a total strain of 20\%, allowing the system to age during at least 5~ns at each temperature. Fig.~\ref{FigEpotTemp}(a) shows the strain dependence of the PED as a function of temperature. For temperatures up to 900~K, the low-strain harmonic regime is again observed, and this is followed by an elastic to plastic transition leading to a steady state. The steady-state PED decreases rapidly with temperature, from $\sim$0.12~eV/atom at 10~K to $\sim$0.04~eV/atom at 300~K, to less than 0.01~eV/atom at 900~K. Increasing temperature is therefore similar to reducing the strain rate.

The results are different for temperatures above 900~K, on which we zoom in in Fig.~\ref{FigEpotTemp}(b): rapid, very small ($\sim$0.01~eV/atom) energy fluctuations, but no saturation, are observed: clear plastic deformations can no longer be defined and shear effects are completely compensated by thermal relaxation.

Interestingly, the steady-state PED curves are all nearly parallel, a behavior also observed by Rottler~\eal~\cite{Rottler} on stress-strain curves at different temperatures. At low temperatures, the potential energy increase is essentially controlled by the shear-imposed displacements and we observe almost no self-diffusion; at high temperature, in contrast, our results reveal a competition between the stress imposed by shear deformations and temperature-enhanced self-diffusion that favors annealing.

\section{\label{structure}Structural modifications during shear deformations}

We discuss in this section the microscopic changes that take place under shear, focussing on the radial distribution function (RDF), $g(r)$, that provides a global picture of structural changes, and local coordination, which we can relate to solid-like or liquid-like behavior.

\subsection{Radial distribution functions}

\begin{figure}[h!]
\includegraphics[width=0.92\figurewidth]{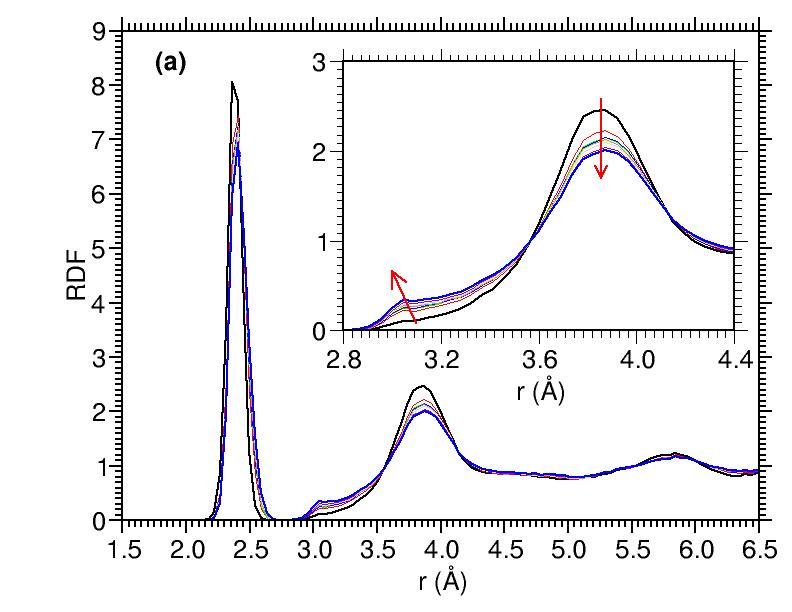}
\includegraphics[width=0.92\figurewidth]{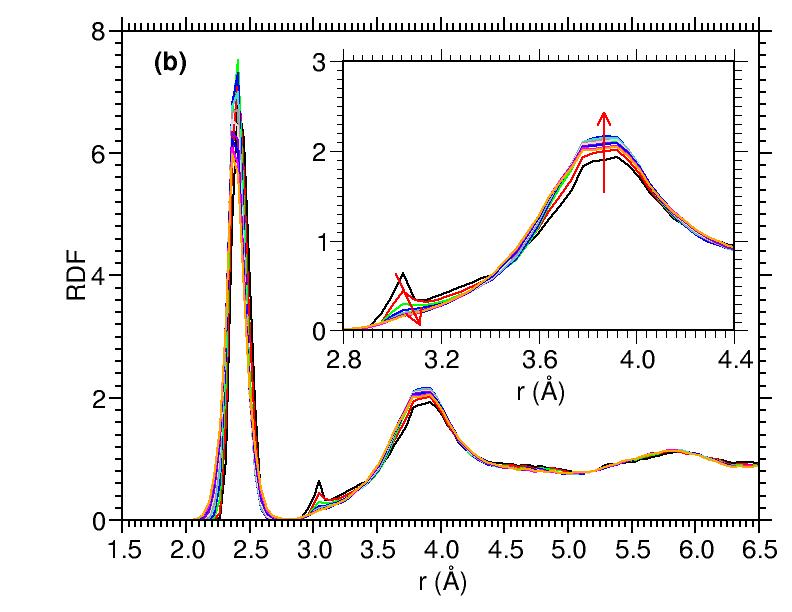}

\caption{Radial distribution function (a) as a function of shear velocity at 300~K, and (b) as a function of temperature at $v_s = 10^{-3}$~\Vshear. The arrows indicate the direction of increasing shear velocity or temperature (see text).}

\label{FigRdf}
\end{figure}

Figure~\ref{FigRdf} (a) shows the average RDF computed for different strain rates at 300~K. To correct for boundary effects, the local, atom-specific RDF is normalized by the fraction of the surface of the sphere surrounding it, $4\pi\,r^{2}$, that fits into the system. Thus, the normalization factor depends on the distance of the particle from the walls in the $y$ direction. Because deformations in the elastic regime do not lead to permanent damage, we focus here on the plastic steady-state regime and average over the last 200 configurations at the maximum strain of 20\%.

Overall, very little difference between the various RDFs is observed: the positions of the first and second peaks are essentially unchanged and match those of the annealed model. While the width of the first-neighbor peak increases slightly, most changes take place near the second-neighbor peak (cf.\ inset): as the height of this peak decreases, a new structure appears at shorter distances ($\sim$2.8~\Ang), which develops as the shear velocity is increased. This feature is related to the appearance of 5-fold coordinated atoms, as already shown in Refs.~\cite{Argon, Demkowicz}: plastic deformations are associated with the presence of higher-coordination liquid-like particles, so that the coordination number increases with shear velocity, which confirms the tendency of mechanical shear deformations to increase disorder.

At fixed shear velocity ($v_s = 10^{-3}$~\Vshear), the new structure formed by shear deformation is more pronounced at lower temperatures, as one can see from the inset to Fig.~\ref{FigRdf} (b) where we present the RDFs at different temperatures. As temperature increases, the system can more easily compensate for the applied strain, which leads to a less pronounced structure between the first and second peak, until it vanishes completely at high temperature, thus confirming that the action of shear dimishes as temperature increases. The fact that the amplitude of the new structure increases with shear velocity and decreases with temperature signals, again, the existence of a competition between shear velocity and temperature as to their effects on the structure.

\subsection{\label{defectssec}Coordination number and defects fraction}

We examine now the evolution of the number of coordination defects under shear, counting atoms within a cutoff of 2.8~\Ang\ which corresponds to the minimum between the first and second neighbor peaks in the annealed-cell RDF. We focus on 5-fold-coordinated atoms since the number of other coordination defects remains very small during the whole process: the initial system contains $\sim$97\% of perfectly coordinated atoms, less than 3\% of 5-fold and only 0.06\% of 3-fold-coordinated atoms.

\begin{figure}[h!]
\includegraphics[width=0.92\figurewidth]{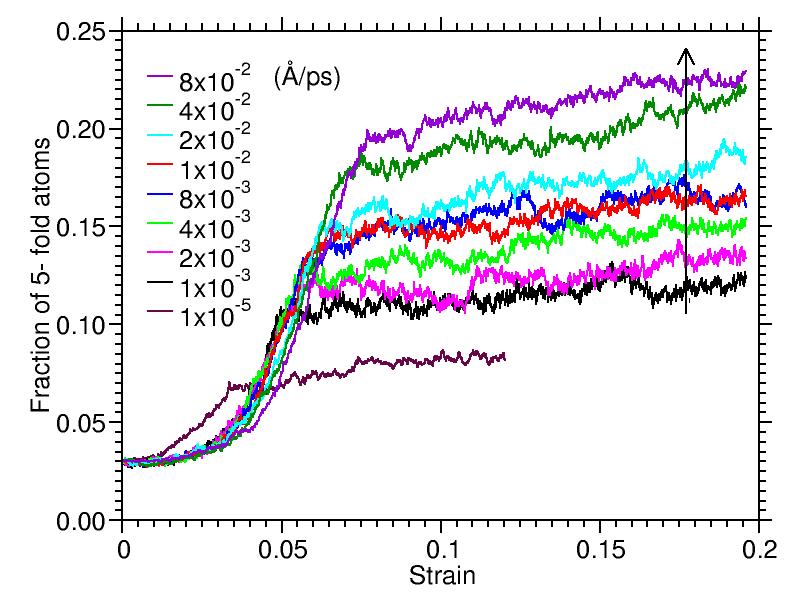}

\caption{Fraction of 5-fold coordinated atoms as a function of strain. The different curves correspond to shear velocities, as indicated. The arrow indicates the direction of increasing shear velocity.}
\label{FigCordTimeShear300}
\end{figure}

The evolution with strain of the population of 5-fold-coordinated defects is displayed in Fig.~\ref{FigCordTimeShear300} for the various shear velocities considered, at 300~K. While the behavior of these curves is similar to that of the PED and the pressure presented earlier, with well-defined elastic and plastic regimes, it differs in two notable ways: First, there is no decrease in the number of defects at the yield stress, but only a sharp flattening of the curve. This suggests that, although the concentration of defects definitely increases in the elastic regime, there is no significant structural reorganization. At the elastic to plastic transition, even though some energetic relaxation takes place at the onset of plasticity, the defects created do not anneal. Second, the population of defects for the smallest shear velocity ($10^{-5}$~\Vshear) evolves along three regimes: a very small elastic region for strains below 0.01, followed by a steady increase of the concentration of defects associated with a fall of equilibrium, and finally a transition to the steady-state plastic regime at a strain of $\sim$0.03. For slow enough shears, therefore, it appears as though, in contrast to crystalline systems, the perfectly elastic regime disappears: amorphous silicon shows a continuous distribution of energy barriers leading to structural rearrangements, and any amount of shear can move the system from one minimum to another~\cite{Valiquette, Kallel10}.

Shear affects the structure very significantly. For the highest shear velocity considered, $8 \times 10^{-2}$~\Vshear, the proportion of 5-fold-coordinated defects increases from as little as 3\% initially to 23\% in the steady state regime; it falls to about 12\% upon decreasing the strain rate by a factor of 80 ($v_s=10^{-3}$~\Vshear), and is only $\sim$8\% for the very lowest shear velocity investigated ($10^{-5}$~\Vshear). Thus, even at 300~K, the system manages to anneal itself. Even for the highest defect levels, other defects (e.g., 3-fold and 6-fold coordinated atoms) remain rare, well below 0.5\%. Interestingly, these results, obtained using the EDIP potential, are in good agreement with those obtained by Talati \eal~ using the Stillinger-Weber potential~\cite{Talati}.

\subsection{Localization of plastic deformations: defect distribution}

The propagation of plastic deformations can be studied by monitoring the proportion of defects --- here 5-fold coordinated atoms --- throughout the cell. To this end, we divide the system into 12 layers along the $y$-direction and calculate the concentration of defects as a function of strain; the first and last layers (the walls) are not considered since they are used for applying the strain.

\begin{figure}[h!]
\includegraphics[width=0.5\figurewidth]{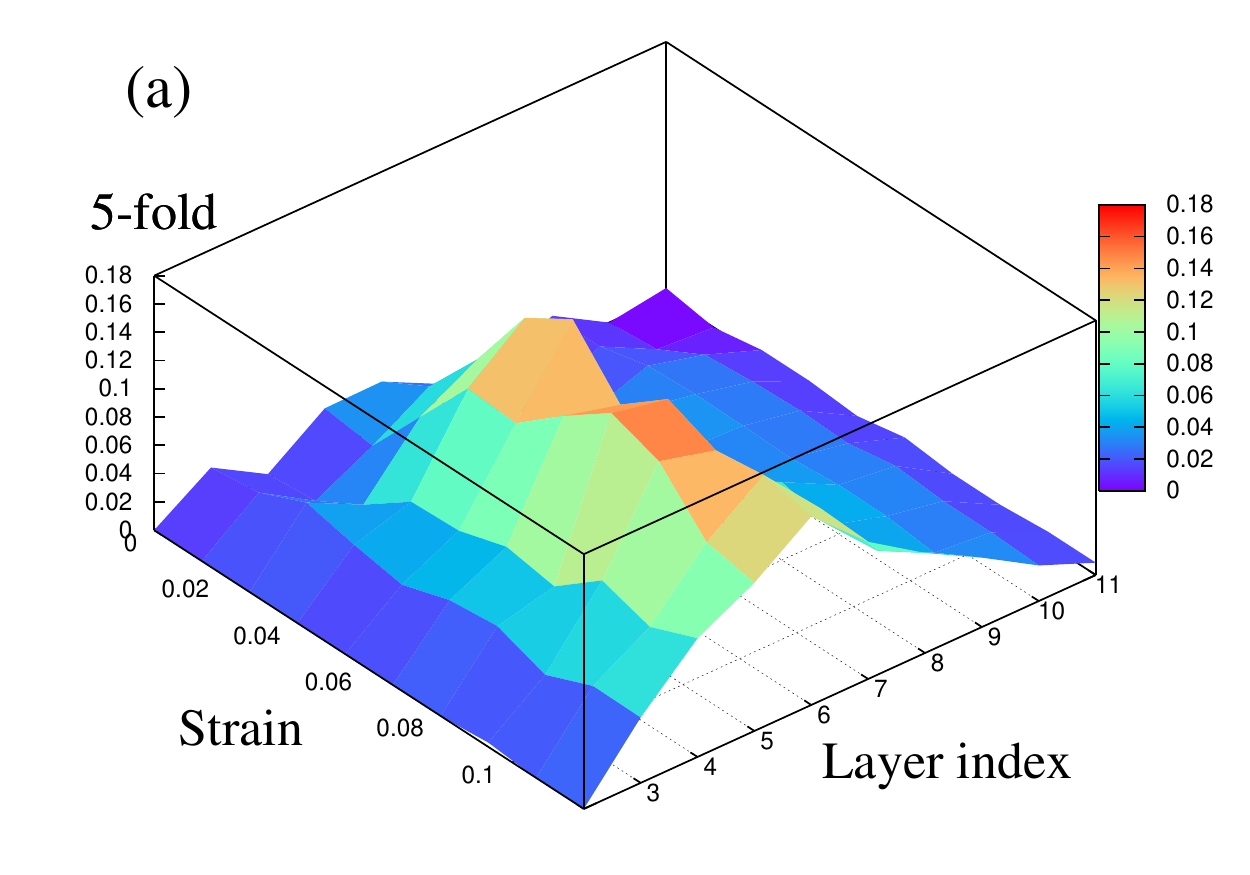}
\includegraphics[width=0.5\figurewidth]{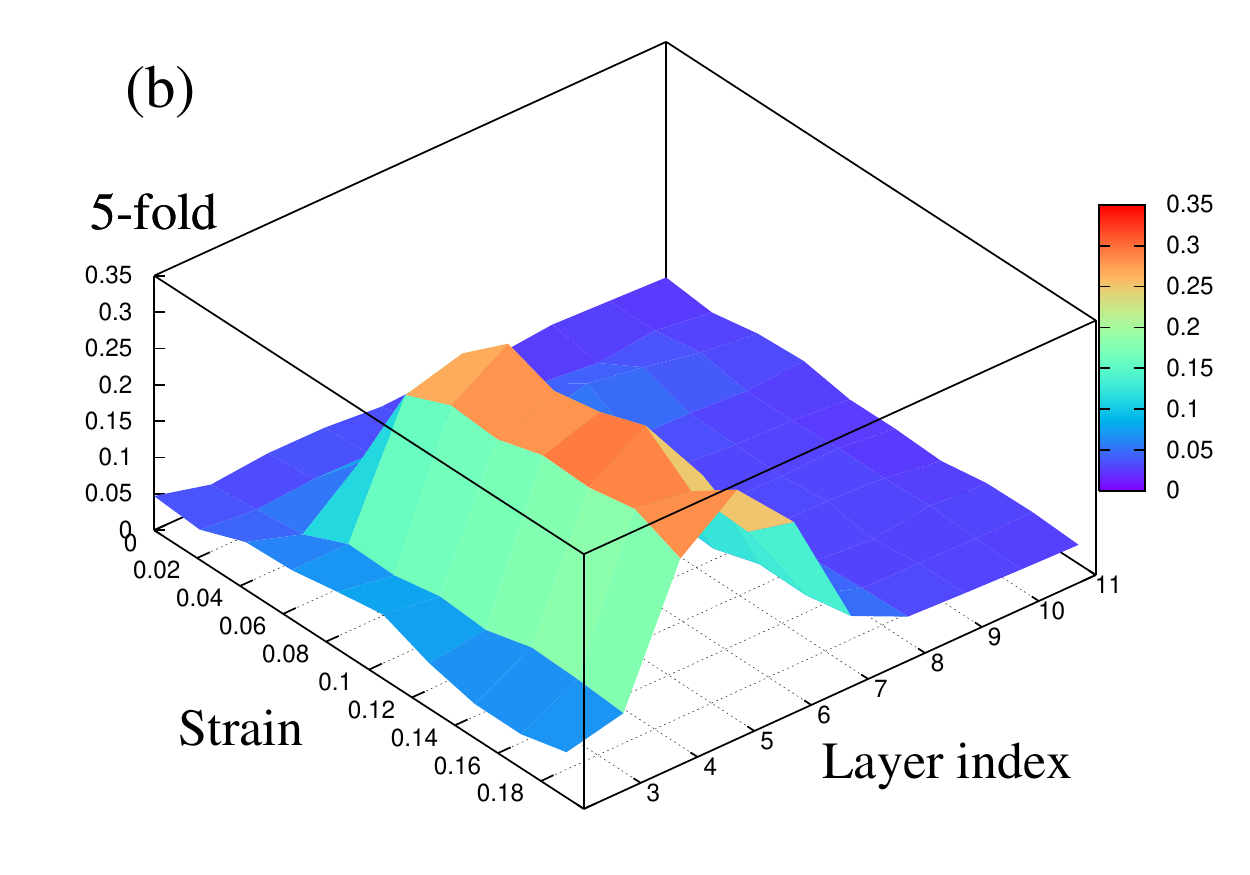}
\includegraphics[width=0.5\figurewidth]{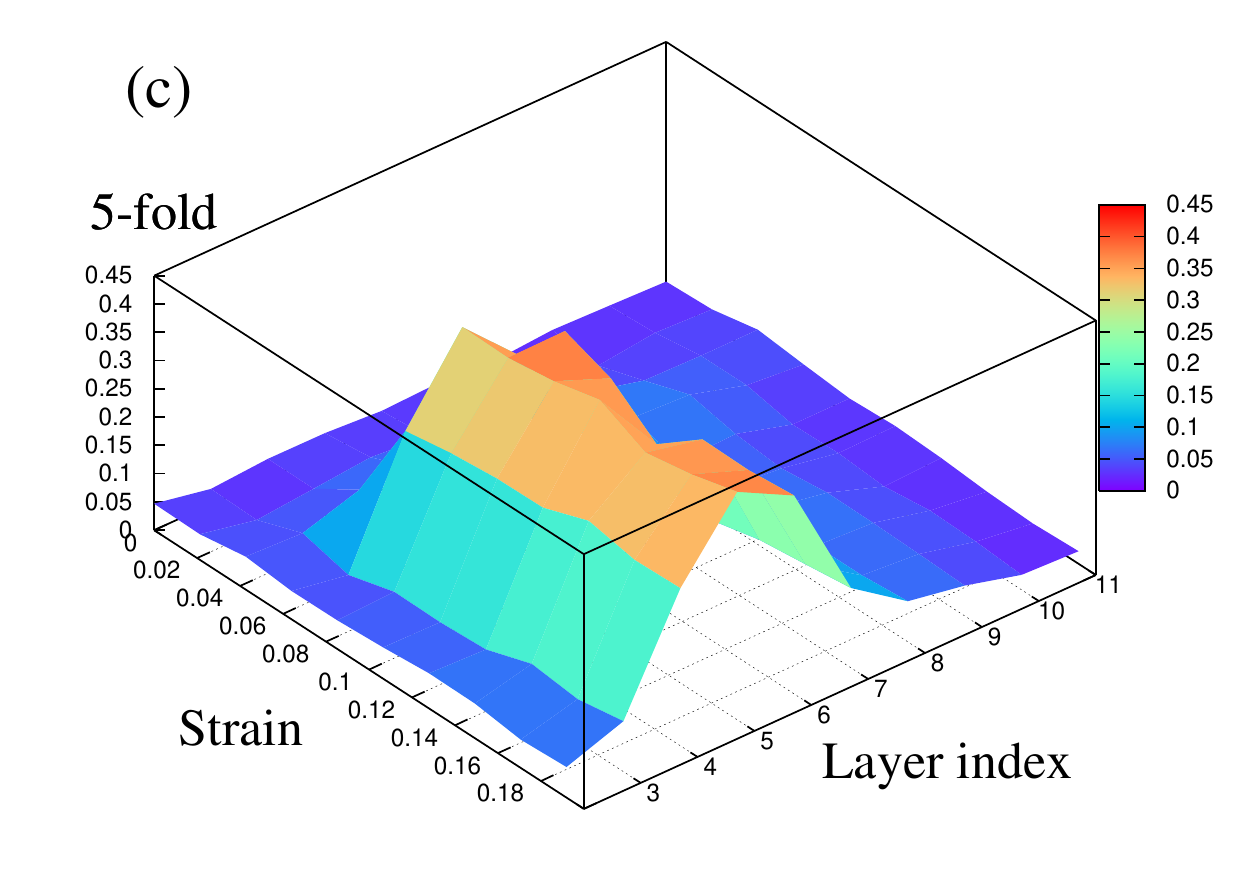}
\includegraphics[width=0.5\figurewidth]{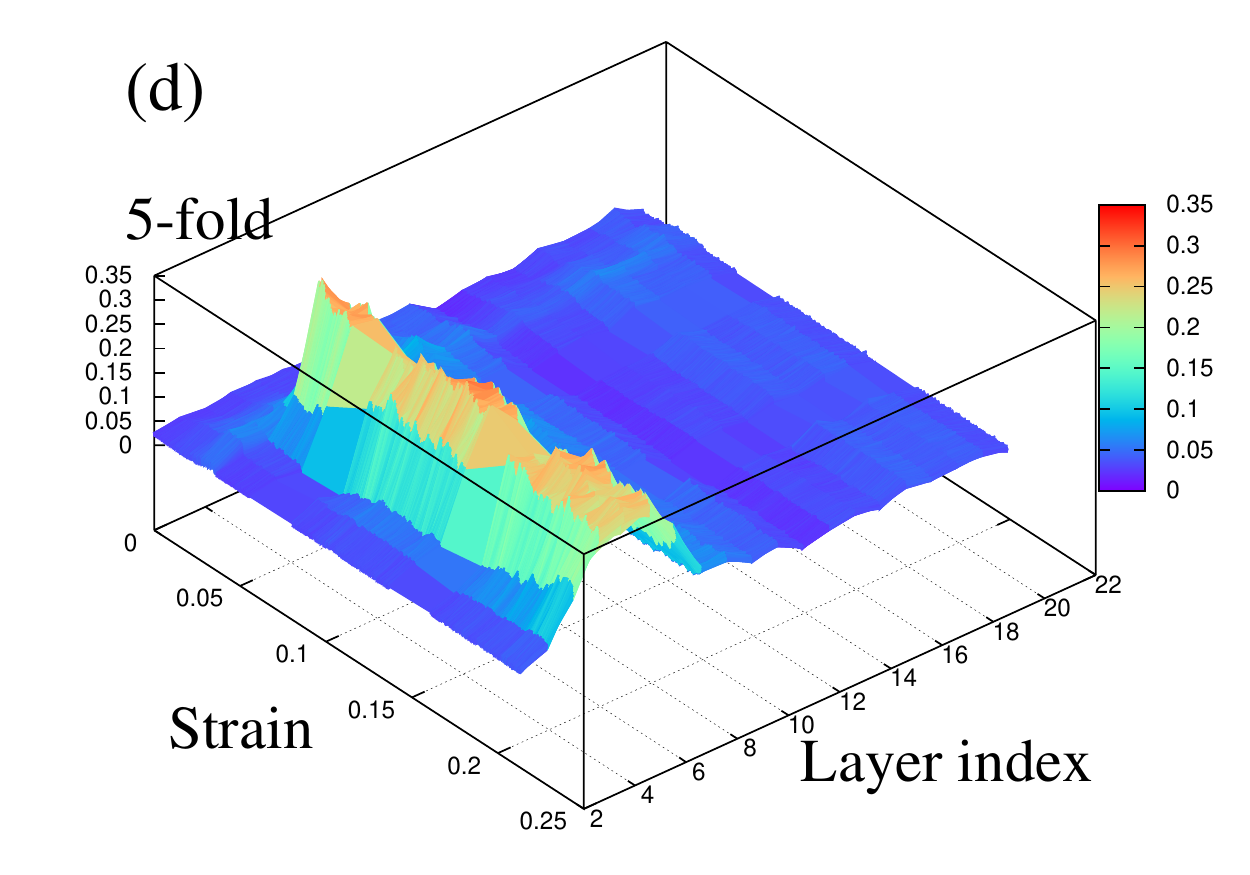}

\caption{Fraction of 5-fold coordinated atoms at 300~K as a function of strain and layer index; (a), (b) and (c) correspond to shear velocities (a) $10^{-5}$~\Vshear, (b) $10^{-3}$~\Vshear, and (c) $10^{-1}$~\Vshear, at 300~K. The configurations are divided into 12 layers along the $y$-direction; layers 1 and 12 are not shown as they correspond to the frozen walls. The figure (d) correspond to a system of 16000 particles and a shear velocity of  $5 \times 10^{-4}$~\Vshear. The number of layers is 22.}
\label{FigSnap5CordVs}
\end{figure}

The results are shown in Figure~\ref{FigSnap5CordVs} for three different shear velocities at 300~K. For all cases considered we find that, at low strain, in the elastic regime, the distribution of defects is low but uniform across the system; at the elastic-to-plastic transition, however, strong inhomogeneities appear along the $y$-direction. At high shear velocities, $10^{-1}$~\Vshear~, Fig.~\ref{FigSnap5CordVs} (c), 5-fold defects form near the center of the system and reach a very high concentration ($>$30\%) in the steady-state regime: plastic deformations are highly localized in these central layers and the 3 to 4 layers near the walls remain largely unaffected. At smaller shear velocities, plastic deformations propagate through a wider portion of the cell, reaching the full width of it for $v_s=10^{-5}$~\Vshear~[Fig.~\ref{FigSnap5CordVs} (a)]. For the latter case, at low strain, defects first appear in the layers closest to the walls (layers (2 and 11) before propagating across the system. In the plastic regime, the defect density increases almost linearly with the distance to the walls, with all layers affected by plastic deformations.

A direct comparison of the figures [Figs.~\ref{FigSnap5CordVs} (b, c and d)] or by computing the distribution of the defects fraction across the y-directions, we found that the width of the region affected by shear deformations -- where the the defects fraction increases significantly -- is almost the same for two different systems. According to these observations, the localization of the shear deformations in a few layers is not size dependent at least for systems larger than 8000 particles.

\begin{figure}[h!]
\includegraphics[width=0.5\figurewidth]{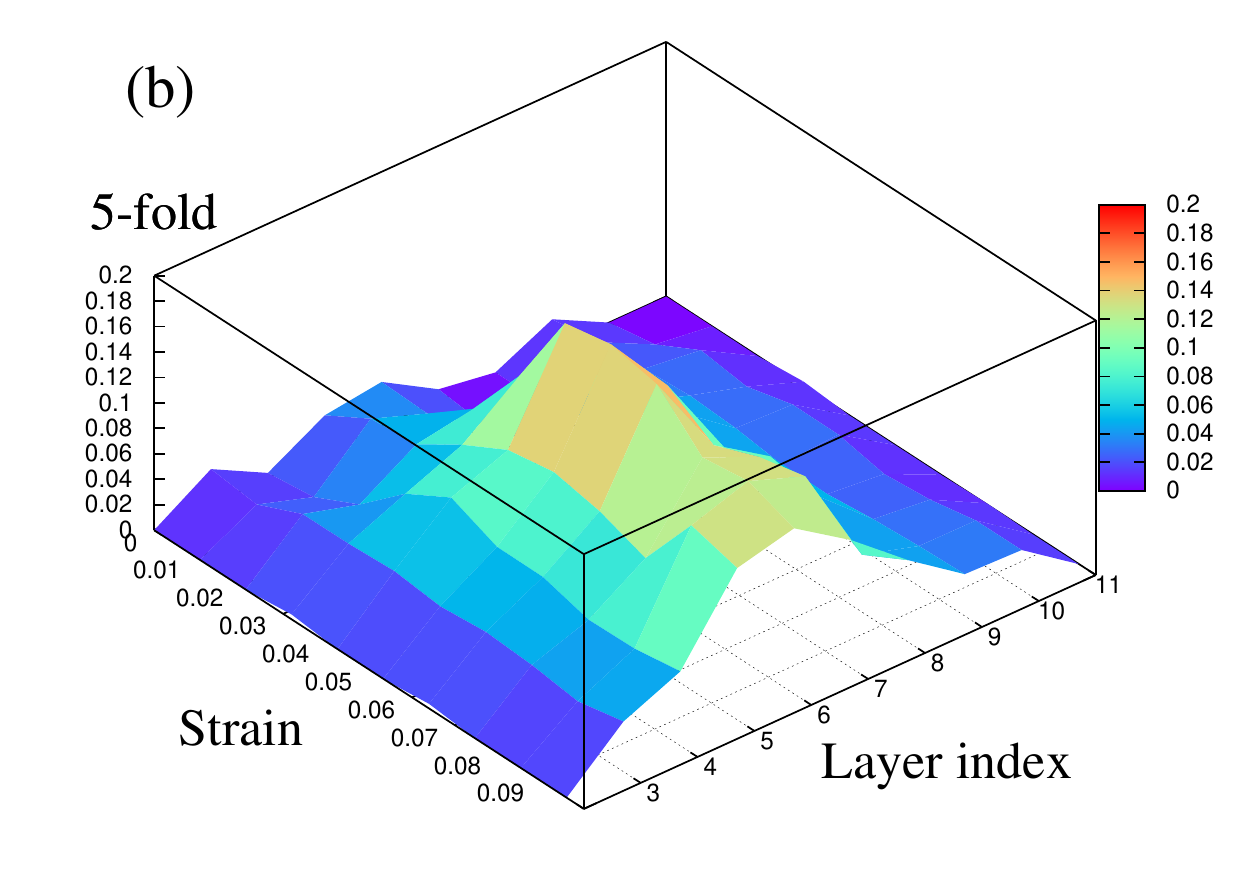}
\includegraphics[width=0.5\figurewidth]{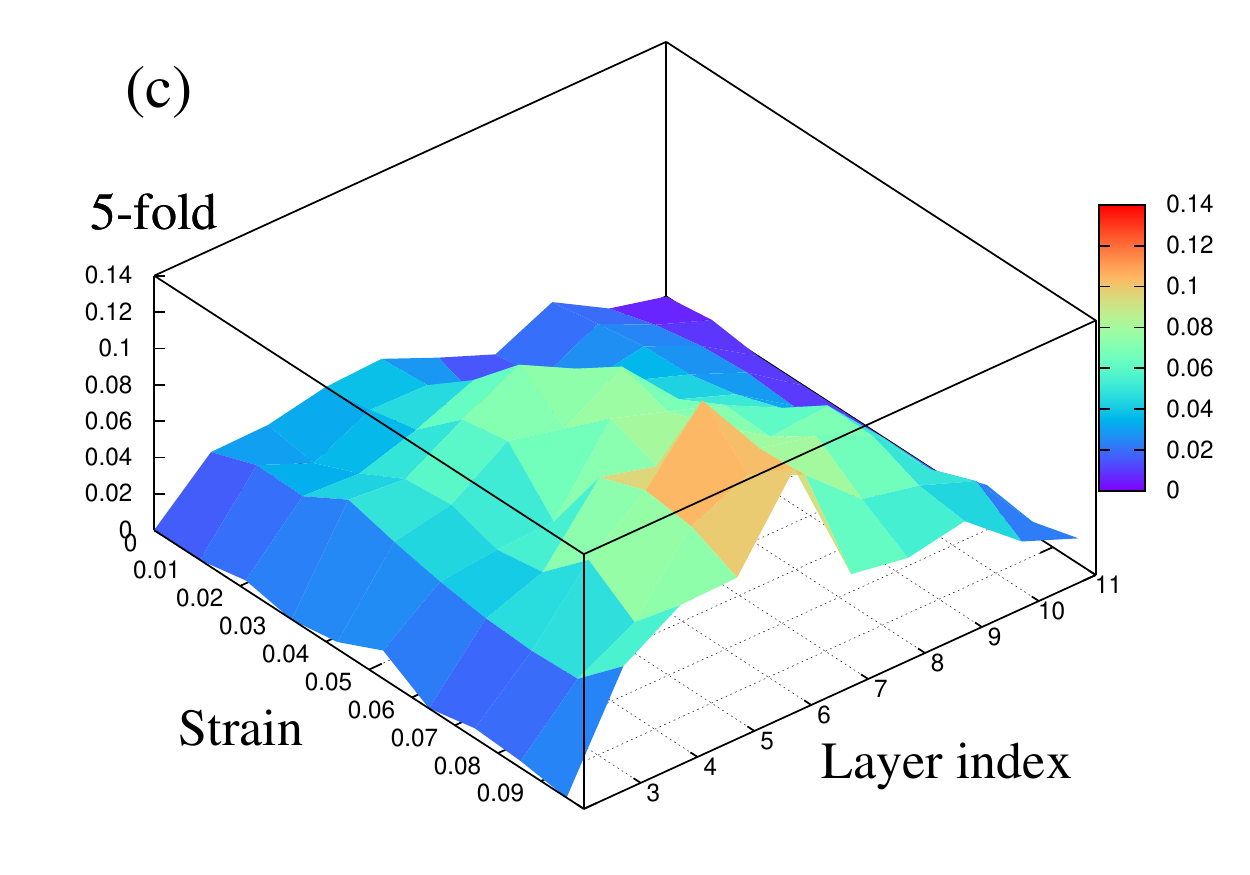}
\includegraphics[width=0.5\figurewidth]{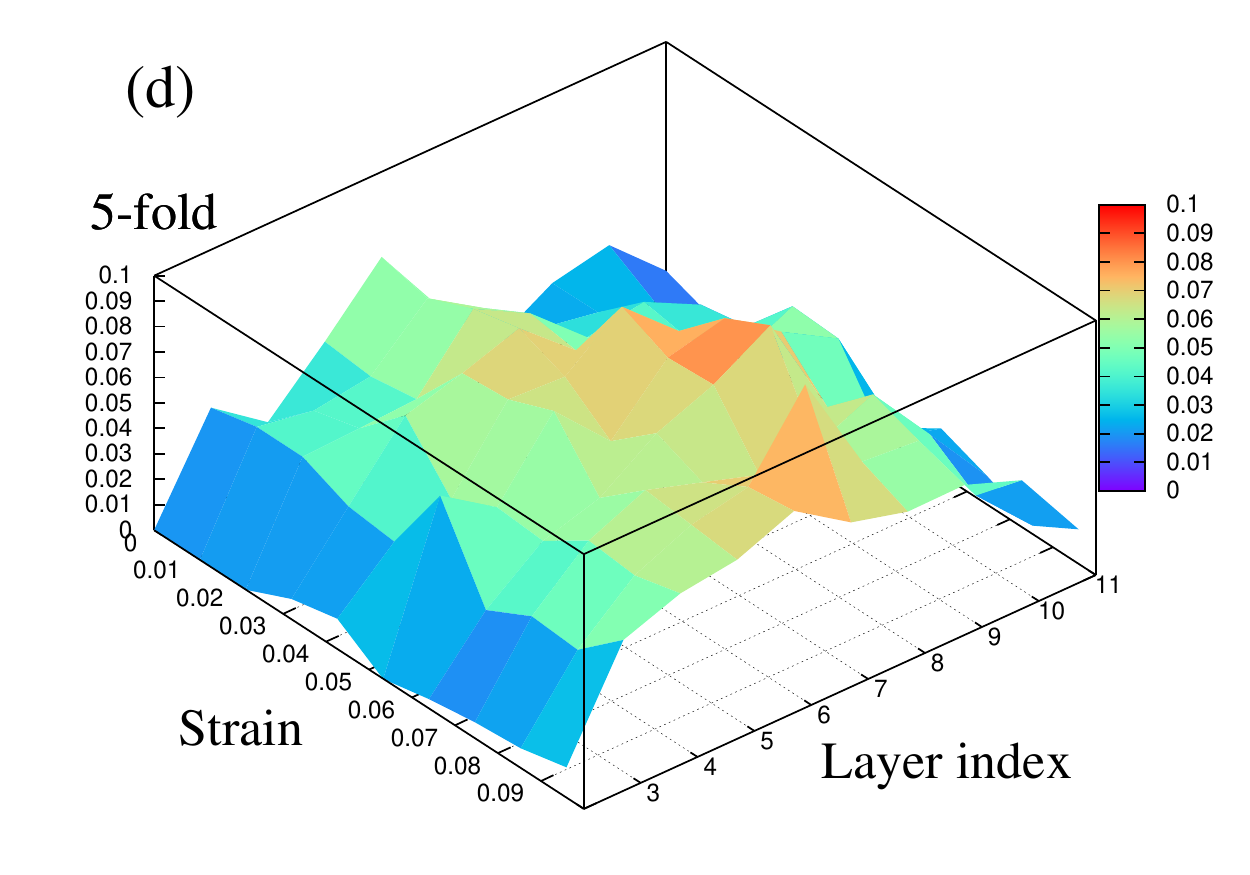}

\caption{Fraction of 5-fold coordinated atoms as a function of strain and layer index, at (a) 500~K, (b) 700~K, and (c) 900~K, for a shear velocity of $10^{-3}$~\Vshear.}
\label{FigSnap5CordTemp}
\end{figure}

The effect of temperature can be assessed in Fig.~\ref{FigSnap5CordTemp} where we plot the corresponding results at temperatures of 500~K, 700~K and 900~K, using an intermediate shear velocity $10^{-3}$~\Vshear. While defects are localized in the few layers near the middle of the structure at 300~K [Fig. \ref{FigSnap5CordVs} (c)], the distribution broadens with temperature: all layers are affected by the shear at 700~K and the distribution becomes almost flat at 900~K, even at small strain.

\section{\label{discussion}Discussion}

In agreement with the results observed on stress-strain curves of the elastic and plastic deformations of different materials~\cite{Falk04, Ediger, Liu93, Weeks, Argon, Demkowicz, Falk, Maloney, Ivashchenko, Talati}, the system's response to strain can divided into three regimes. At low strain, the system responds with strain-independent elastic deformations, as observed in the PED and the pressure, for example. Although the number of local defects, here over-coordinated atoms, increases during this phase, a large fraction of these are reversible. The second regime, at intermediate strain, depends on the shear velocity and is associated with the elastic to plastic transition. Here, the absolute values of the PED and the pressure reach a maximum, then fall off to smaller values following a cascade of bond rearrangements that reduces the strain buildup, the drop being larger for faster shear rates and lower temperatures. Surprisingly, however, this is not accompanied by a similar drop in the number of defects. As the energy is released, the defects are therefore stabilized by irreversible atomic repositioning. The third regime corresponds to steady-state plastic deformations and occurs at high strain values. In this regime, the various microscopic and thermodynamic quantities are essentially strain-independent. For the pressure, the plastic regime rapidly reaches a plateau after the break-down. For the potential energy, the convergence is somewhat slower and the steady-state value is only slightly lower than the maximum at the elastic-to-plastic transition. The curves are also shifted to lower energies with decreasing shear velocity or increasing temperature~\cite{Rottler}. This contrasts with the microscopic density of 5-fold defects, which shows a steady increase and is not yet converged for most values of the shear velocity and temperature at a strain of 0.2.

Our results on the effects of shear and temperature allow us to better characterize the spatial inhomogeneities involved in the response to shear. While defects form almost uniformly across the atomic layers in the elastic phase, plastic deformations are concentrated within a narrow region near the center of the box at low temperatures and high shear rates, and the fraction of liquid-like 5-fold coordinated atoms in these layers reaches almost 40\%, in line with the results of Refs.~\cite{Argon, Demkowicz}. Indeed, the cell is breaking into two parts with an almost fluid interface between them. Such a sharp break is costly however and is imposed by kinetic considerations: the relaxation time is too short to permit a global response of the system to the external perturbation. The system is, in effect, frozen, and stress can build up significantly, which can only be released by a size-wide rearrangement, creating the observed sharp break. This inhomogeneity explains why the radial distribution function is only slightly affected by an average defect concentration of more than 20\%.

As the temperature rises or the shear velocity decreases, the relaxation time becomes comparable with the shear rate and the response to perturbation is spread to larger and larger regions, preventing a large build-up characterized by a significant increase in potential energy or defect concentration. And while 5-fold coordinated atoms remain the main type of defects by which this response takes place, it is no longer possible to speak of localized centers of deformation, since the whole system is affected in the plastic regime.

\begin{figure}[h!]
\includegraphics[width=0.92\figurewidth]{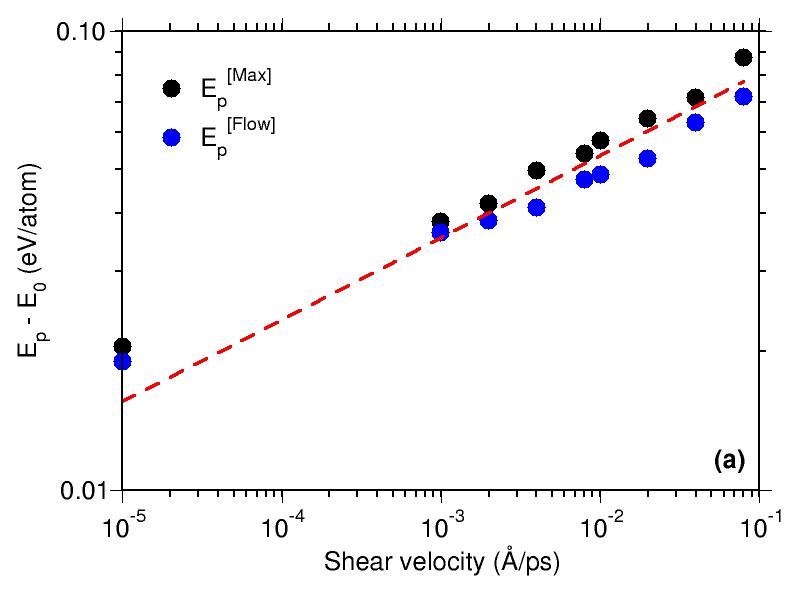}
\includegraphics[width=0.92\figurewidth]{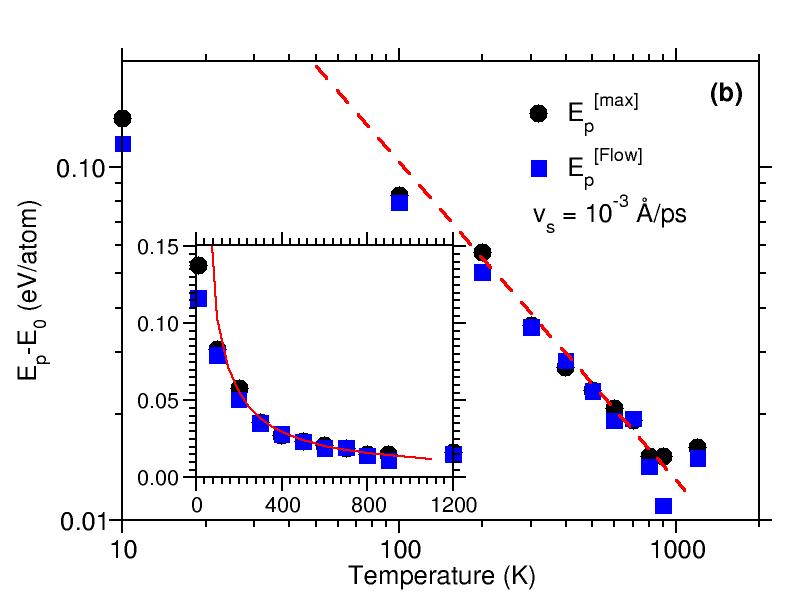}

\caption{Log-log plot of the potential energy (maximum and mean value in the steady state for high strain values) as a function of (a) shear velocity at fixed temperature (300~K), and (b) temperature at fixed shear velocity ($10^{-5}$~\Vshear). The inset on the panel (b) shows the same curve in a linear scale.}
\label{FigLogLog}
\end{figure}

The relation between temperature and shear velocity can be seen in Fig.~\ref{FigLogLog} which shows the evolution of the maximum PED and the mean value of the PED in the plastic deformation regime as a function of shear velocity and temperature; the data are extracted from Figs.~\ref{FigEpotPressShear300} and~\ref{FigEpotTemp}. For the PED, we observe almost the same power-law relation between the shear-velocity and the energy at the yield-stress and in the steady state plastic regime, in both cases with an exponent of $\simeq 0.18$ in a wide range of shear velocities between $10^{-3}$~\Vshear~and $8 \times 10^{-2}$~\Vshear. For the smallest shear velocity, the power-law relation seems to be less reliable and the effects of shear appear to be more important than could be expected. Since thermal effects are included in the reference potential energy, this suggests that there is no critical rate for which thermal energy completely obliterates the effects of shearing even though plasticity will spread to always larger fractions of the system.

In the bottom panel of Fig.~\ref{FigLogLog}, a similar power-law behavior is observed for temperatures between 150~K and 900~K, with an exponent of $-$0.9. Again, even at high temperature (but below melting), the effects of shear are not fully compensated but are strongly reduced. At low temperature, the power-law relation breaks down because the bonding energy introduces a hard threshold on the amount of strain that can be stored in the system.

The origin of these power-law relations can be related to the energy landscape structure of the system. Because the activation energy necessary to cross a barrier is uncorrelated with the energy difference between the top of the barrier and the final minimum, only the shape of the forward activation-energy barrier distribution is important~\cite{Kallel10}. From numerical calculations~\cite{Kallel10}, it is known that this activation-energy barrier distribution at $T=0$~K is a continuous function that can be fitted to:
\begin{equation}
G_{FB}(E) = A E \exp\left\{ - \frac{(E-\langle E_{FB}^{rel} \rangle)^2}{2 \sigma_{FB}^2}\right\},
\label{eq1}
\end{equation}
where $E$ is the barrier energy, $A$ a normalisation factor, and $E_{FB}^{rel}$ and $\sigma_{FB}$ two parameters that depend on the system and the degree of relaxation. At finite temperature, for a well-relaxed sample, barriers below $k_BT$ are already \emph{consumed}. The shearing here effectively increases the system's energy, thus decreasing the height of available barriers, and, from there, increases the crossing probability at a given temperature. Because energy barriers of any height may exist, it
is possible for the system to accommodate shearing at least partially, at any temperature, as long as the shearing rate is slow with respect to the attempt-to-jump frequency.

\section{\label{conclude}Concluding remarks}

We have investigated the properties of amorphous silicon subject to external mechanical shear deformations using classical MD simulations with the Environment Dependent Inter-atomic Potential (EDIP). The shear deformations are introduced by moving one wall in the shear direction~\cite{Mokshin}. We find that both energetic and structural properties of the system depend strongly on the shear velocity and the imposed strain, as well as the temperature at which the deformations are applied. At low temperatures and for all shear velocities investigated, we observe a systematic increase in disorder associated with an increase in the fraction of coordination defects, in agreement with the results of the references~\cite{Talati, Rottler}. Interestingly, the shear-induced energy can be written in terms of a power-law of both temperature and shear velocity: $\Delta E \propto v_s^{\alpha} T^{-\beta}$, with $\alpha \simeq 0.18$ and $\beta \simeq 0.9$. Finally, we observe a very strong dependence of the spatial distribution of defects on these two quantities. For low temperatures or high shear velocities, defects are localized in a narrow region in the middle of the cell. As the temperature increases or the shear rate slows down, this region becomes wider until it covers the whole system. This effect is due to the existence of a continuous distribution of activation energy barriers that allows the system to relax at any temperature provided that it has ample time to do so, as determined by the shear rate.

\begin{acknowledgments}

This work is supported in part by the FQRNT (Qu\'ebec), NSERC (Canada) and the Canada Research Chair foundation. The calculations were carried out on the computers of the ``R\'eseau Qu\'eb\'ecois de Calcul de Haute Performance (RQCHP)'' whose support is gratefully acknowledged.

\end{acknowledgments}


\begin{thebibliography}{5}

\bibitem {WWW} {F. Wooten, K. Winer, and D. Weaire, Phys. Rev. Lett. {\bf 54}, 1392 (1985).}

\bibitem {Mousseau98} {N. Mousseau and G. T. Barkema, Phys. Rev. E {\bf 57}, 2419 (1998).}

\bibitem {Barkema98} {G. T. Barkema and N. Mousseau, Phys. Rev. Lett. {\bf 81}, 1865 (1998).}

\bibitem {Balamane} {H. Balamane, T. Halicioglu, and W.A. Tiller, Phys. Rev. B {\bf 46}, 2250 (1992).}

\bibitem {Bazant} {M.Z. Bazant and E. Kaxiras, Phys. Rev. Lett. {\bf 77}, 4370 (1996); M.Z. Bazant, E. Kaxiras, J.F. Justo, Phys. Rev. B {\bf 56}, 8542 (1997).}

\bibitem {Justo} {J.F. Justo, M.Z. Bazant, E. Kaxiras, V.V. Bulatov, and S. Yip, Phys. Rev. B {\bf 58}, 2539 (1998).}

\bibitem {SW} {F.H. Stillinger and T.A. Weber, Phys. Rev. B {\bf 31}, 5262 (1985).}

\bibitem {Tersoff} {J. Tersoff, Phys. Rev. B {\bf 37}, 6991 (1988).}

\bibitem {Gillespie} {B.A. Gillespie, X.W. Zhou, D.A. Murdick, H.N.G. Wadley, R. Drautz, and D.G. Pettifor, Phys. Rev. B {\bf 75}, 155207 (2007).}

\bibitem {Kluge} {M.D. Kluge, J.R. Ray, and A. Rahman, Phys. Rev. B {\bf 36}, 4234 (1987).}

\bibitem {Biswas} {R. Biswas, G.S. Grest, and C.M. Soukoulis, Phys. Rev. B {\bf 36}, 7437 (1987).}

\bibitem {Valiquette} {F. Valiquette and N. Mousseau, Phys. Rev. B \textbf{68}, 125209 (2003).}

\bibitem {Stich} {I. Stich, R. Car, M. Parrinello, and S. Baroni, Phys. Rev. B {\bf 39}, 4997 (1989).}

\bibitem {Car} {R. Car and M. Parrinello, Phys. Rev. Lett. {\bf 60}, 204 (1988).}

\bibitem {Bernstein06} {N. Bernstein, J.L. Feldman, and M. Fornari, Phys. Rev. B {\bf 74}, 205202 (2006).}

\bibitem {Makhov} {D.V. Makhov and L.J. Lewis, Phys. Rev. Lett. {\bf 92}, 255504 (2004).}

\bibitem {Feldman} {J.L. Feldman, N. Bernstein, D.A. Papaconstantopoulos, and M.J. Mehl, Phys. Rev. B {\bf 70}, 165201 (2004).}

\bibitem {Huang} {H. Shi-Ping and W. Wen-Chuan, Chin. Phys. Lett. {\bf 21}, 2482 (2004).}

\bibitem {Drabold} {P. Biswas and D. A. Drabold, J. Non. Cryst. Sol. \bf{354}, 2697 (2008).}

\bibitem {Mousseau00} {N. Mousseau and G. T. Barkema, Phys. Rev. B {\bf 61}, 1898-1906 (2000).}

\bibitem {David} {D. Redfield and R.H. Bube, Phys. Rev. Lett. {\bf 65}, 464 (1990).}

\bibitem {Roorda} {S. Roorda, W.C. Sinke, J.M. Poate, D.C. Jacobson, S. Dierker, B.S. Dennis, D.J. Eaglesham, F. Spaepen, and P. Fuoss, Phys. Rev. B {\bf 44}, 3702 (1991).}

\bibitem {Coffa} {S. Coffa, F. Priolo, and A. Battaglia, Phys. Rev. Lett. {\bf 70}, 3756 (1993).}

\bibitem {Raymond} {R. Lutz and L.J. Lewis, Phys. Rev. {\bf 47}, 9896 (1993).}

\bibitem {Knief} {S. Knief, W. von Niessen, and T. Koslowski, Phys. Rev. B {\bf 58}, 4459 (1998).}

\bibitem {Urli} {X. Urli, C.L. Dias, L.J. Lewis, and S. Roorda, Phys. Rev. B {\bf 77}, 155204 (2008).}

\bibitem {Peressi} {M. Peressi, M. Fornari, S. Degironcoli, L. Desantis, and A. Baldereschi, Phil. Mag. B {\bf 50}, 515 (2000).}

\bibitem {Goedecker} {S. Goedecker, T. Deutsch, and L. Billard, Phys. Rev. Lett. {\bf 88}, 235501 (2002).}

\bibitem {Amarendra} {G. Amarendra, R. Rajaraman, G. Venugopal Rao, K.G.M. Nair, B. Viswanathan, R. Suzuki, T. Ohdaira, and T. Mikado, Phys. Rev. B {\bf 63}, 224112 (2001).}

\bibitem {Pantelides} {S.T. Pantelides, Phys. Rev. Lett. {\bf 57}, 2979 (1986).}

\bibitem {Kim} {E. Kim, Y.H. Lee, Phys. Rev. B {\bf 51}, 5429 (1995); E. Kim, Y.H. Lee, C. Chen, and T. Pang, Phys. Rev. B {\bf 59}, 2713 (1999).}

\bibitem{zhang08} {Zhang, Y. Pan, F. Inam, and D. A. Drabold, Phys. Rev B \textbf{78}, 195208 (2008).}

\bibitem {Drabold10} {D.A. Drabold, Eur. Phys. J. B \textbf{68}, 1 (2009).}

\bibitem {Rice} {J.R. Rice, J. Mech. Phys. Solids {\bf 40}, 239 (1992).}

\bibitem {Acharya} {A. Acharya, J. Mech. Phys. Sol. \textbf {49}, 761 (2001).}

\bibitem{Gaucherin}{G. Gaucherin, F. Hofmann, J. P. Belnoue and A. M. Korsunsky, Procedia Engineering, \textbf{1}, (2009) 241-244.}

\bibitem{Mesarovic}{S. Dj. Mesarovic, R. Baskaran and A. Panchenko, J. Mech. Phys. Sol., \textbf{58}, (2009) 311-329.}

\bibitem{Bonifaz}{E.A. Bonifaz and N.L. Richards, International Journal of Plasticity, \textbf{24}, (2008) 289-301.}

\bibitem{Devincre}{B. Devincre and L. Kubin, Comptes Rendus Physique, \textbf{11}, (2010) 274--284.}

\bibitem{Enikeev}{N. A. Enikeev, H. S. Kim, and I. V. Alexandrov, Materials Science and Engineering: A, \textbf{460--461}, (2007), 619--623.}

\bibitem{Hwang}{H. Wang, K.C. Hwang, Y. Huang, P.D. Wu, B. Liu, G. Ravichandran, C.-S. Han and H. Gao, International Journal of Plasticity, \bf{23}, (2007), 1540--1554.}

\bibitem{Gao}{H. Gao, Y. Huang, W. D. Nix and J. W. Hutchinson, J. Mech. Phys. Sol. \textbf{47} (1999), pp. 1239--1263.}

\bibitem{Han}{Chung-Souk Han, Huajian Gao, Yonggang Huang, William D. Nix, Journal of the Mechanics and Physics of Solids, \textbf{53}, (2005), 1204-1222}

\bibitem {Falk04} {M.L. Falk, J.S. Langer, and L. Pechenik, Phys. Rev. E {\bf 70}, 011507 (2004).}

\bibitem {Ediger} {M.D. Ediger, Annu. Rev. Phys. Chem. {\bf 51}, 99 (2000).}

\bibitem {Liu93} {C.H. Liu and S.R. Nagel, Phys. Rev. B {\bf 48}, 15646 (1993).}

\bibitem {Weeks} {E.R. Weeks, J.C. Crocker, A.C. Levitt, A. Schofield, and D.A. Weitz, Science {\bf 287}, 627 (2000); E.R. Weeks and D.A. Weitz, Phys. Rev. Lett. {\bf 89}, 095704 (2002).}

\bibitem {Helder} {A. Helder, S.L. Klaumuzer, and W. Wesch, Nature Material {\bf 3} (2004) 804.}

\bibitem {ArgonSTZ} {A.S. Argon, Acta Metall. {\bf 27}, 47 (1979); A.S. Argon and H. Kuo, Mater. Sci. Eng. {\bf 39}, 101 (1979); A.S. Argon and L.T. Shi, Acta Metall. {\bf 31}, 499 (1983).}

\bibitem {Falk} {M.L. Falk and J.S. Langer, Phys. Rev. E {\bf 57}, 7192 (1998).}

\bibitem{Langer06} {J.S. Langer, Scripta Materialia {\bf 54}, (2006) 375-379.}

\bibitem {Argon} {A.S. Argon and M.J. Demkowicz, Metallurgical and Materials Transactions A {\bf 39} (2008) 1762.}

\bibitem {Demkowicz} {M.J. Demkowicz and A.S. Argon, Phys. Rev. Lett. {\bf 93}, 025505 (2004); Phys. Rev. B {\bf 72}, 245205 (2005); Phys. Rev. B {\bf 72}, 245206 (2005).}

\bibitem {Talati} {M. Talati, T. Albaret, and A. Tanguy, Eur. Phys. Lett. {\bf 86}, 66005 (2009).}

\bibitem {Allred} {C.L. Allred, X. Yuan, M.Z. Bazant, and L.W. Hobbs, Phys. Rev. B {\bf 70}, 134113 (2004).}

\bibitem {LAMMPS} {S.J. Plimpton, J. Comp. Phys. {\bf 117}, 1 (1995); see http://lammps.sandia.gov/}

\bibitem {Ivashchenko} {V.I. Ivashchenko, P.E.A. Turchi, and V.I. Shevchenko, Phys. Rev. B {\bf 75}, 085209 (2007).}

\bibitem {Caturla} {M.-J. Caturla, T. Diaz de la Rubia, L.A. Marques, and G.H. Gilmer, Phys. Rev. B {\bf 54}, 16683 (1996).}

\bibitem {Nord} {J. Nord, K. Nordlund, and J. Keinonen,  Phys. Rev. B {\bf 65}, 165329 (2002).}

\bibitem {Nakhmanson} {S. Nakhmanson and N. Mousseau, J. Phys.: Condens. Matter {\bf 14}, 6627 (2002).}

\bibitem {Pelaz} {L. Pelaz, L.A. Marqu\'es, and J. Barbolla, J. Appl. Phys. {\bf 96}, (2004) 5947.}

\bibitem{Park}{S. H. Park, H. J. Kim, K. H. Kang, J. S. Lee, Y. K. Choi and O. M. Kwon, J. Phys. D: Appl. Phys. ; \bf{38}, 1511 (2005). }

\bibitem {Krzeminski} {C. Krzeminski, Q. Brullin, V. Cunny, E. Lecat, and F. Cleri, J. Appl. Phys. {\bf 101}, 123506 (2007)}.

\bibitem {Donovan} {E.P. Donovan, F. Spaepen, D. Turnbull, J.M. Poate, and D.C. Jacobson, Appl. Phys. Lett. {\bf 42}, 698 (1983).}


\bibitem {Matsumura} {H. Matsumura and H. Tachibana, Appl. Phys. Lett. {\bf 47}, 833 (1985).}

\bibitem {Laaziri} {K. Laaziri, S. Kycia, S. Roorda, M. Chicoine, J.L. Robertson, J. Wang, and S.C. Moss, Phys. Rev. Lett. {\bf 82}, 3460 (1999).}

\bibitem {Barkema} {G.T. Barkema and N. Mousseau, Phys. Rev. B {\bf 62}, 4985 (2000).}

\bibitem {Mokshin} {A.V. Mokshin and J.-L. Barrat, Phys. Rev. E {\bf 77}, 021505 (2008).}

\bibitem {Buta} {D. Buta, M. Asta, and J.J. Hoyt, Phys. Rev. E {\bf 78}, 031605 (2008).}

\bibitem {Brambilla} {L. Brambilla and L. Colombo, Appl. Phys. Lett. {\bf 77}, 2337 (2000).}

\bibitem {Bernstein98} {N. Bernstein, M.J. Aziz, and E. Kaxiras, Phys. Rev. B {\bf 58}, 4579 (1998).}

\bibitem {Mattoni} {A. Mattoni and L. Colombo, Phys. Rev. B {\bf 78}, 075408 (2008).}

\bibitem {Rottler} {J. Rottler and M.O. Robbins, Phys. Rev. E {\bf 68}, 011507 (2003).}

\bibitem {Klaumunzer} {S. Klaumunzer and G. Schumacher, Phys. Rev. Lett. {\bf 51}, 1987 (1983).}

\bibitem {Hou} {M.-D. Hou, S. Klaumunzer, and G. Schumacher, Phys. Rev. B {\bf 41}, 1144 (1990).}

\bibitem {Maloney} {C.E. Maloney and A. Lemaitre, Phys. Rev. E {\bf 74}, 016118 (2006).}

\bibitem{Bulatov} {V.V. Bulatov and A.S. Argon, Modell. Simul. Mater. Sci. Eng. {\bf 2}, 167 (1994); {\bf 2}, 185 (1994); {\bf 2}, 203 (1994).}

\bibitem{Kallel10} {H. Kallel, N. Mousseau and F. Schiettekatte, Phys. Rev. Lett. {\bf 105}, 045503.}

\end{thebibliography}
\end{document}